\documentclass[preprint,showpacs,preprintnumbers,amsmath,amssymb,aps]{revtex4-1}

\usepackage{graphicx}
\usepackage{dcolumn}
\usepackage{bm}
\usepackage{graphicx}
\usepackage{subfigure}
\usepackage{tikz}
\usepackage{siunitx}
\usepackage{float}
\usetikzlibrary{shapes,arrows}
\begin{document}


\title{Numerical analysis of neutrino physics within a high scale supersymmetry model via machine learning}

\author{Ying-Ke Lei$^{1,2}$} \email{leiyingke@mail.itp.ac.cn} 
\author{Chun Liu$^{1,2}$} \email{liuc@mail.itp.ac.cn} 
\author{Zhiqiang Chen$^{3,2}$} \email{chenzhiqiang14@mails.ucas.ac.cn} 

\affiliation{$^1$ CAS Key Lab. of Theor. Phys., Institute of Theoretical Physics, Chinese Academy of Sciences, Beijing 100190, China }
\affiliation{$^2$ School of Physical Sciences, Univ. of Chinese Academy of Sciences, Beijing 100049, China}
\affiliation{$^3$ Research Center for Brain-Inspired Intelligence, Institute of  Automation, Chinese Academy of Sciences (CASIA), Beijing 100190,China}

\date{\today} 

\begin{abstract} 
A machine learning method is applied to analyze lepton mass matrices 
numerically. The matrices were obtained within a framework of high 
scale SUSY and a flavor symmetry, which are too complicated to be 
solved analytically.  In this numerical calculation, the heuristic 
method in machine learning is adopted.  Neutrino masses, mixings, and 
CP violation are obtained.  It is found that neutrinos are normally 
ordered and the favorable effective Majorana mass is about 
$7\times 10^{-3}$ eV.  
\end{abstract} 

\maketitle

\section{Introduction}

The fermion mass pattern is an interesting problem in elementary 
particle physics.  Notably, in the leptonic sector, some unknown 
physical parameters are still under experimental measurements, such as 
the CP-violating phase in neutrino oscillation and mass ordering of 
neutrinos 
\cite{acciarri2016long,Djurcic:2015vqa,abi2018dune,Abe:2015zbg,Chen:2016qcd}.  
They will provide checks for various theoretical models about the 
fermion mass pattern like that in Refs. 
\cite{Guo:2012xb, Hu:2012ei,Zhao:2012wq,Kawamura:2019mrl}.  

In our last work \cite{Lei:2018lln}, we analyzed a theoretical model 
for fermion masses, which involves high scale supersymmetry (SUSY) and 
a flavor symmetry \cite{Liu:2005qic, Liu:2015rs, Liu:2012qua}.  The 
form of mass matrices has been predicted with all the basic parameters 
being in the natural range.  However, there are many parameters in the 
model, as a consequence, the matrices are still too complicated to get 
a full analytical discussion.  We simplified the analysis by taking 
some phase parameters in the mass matrices to be zero.  Although some 
results were obtained, the conclusion might not be generic.  
Nevertheless, it is necessary to perform the analysis without arbitrary 
approximation.  

There are fifteen parameters in our mass matrices, whereas
experimentally known quantities are the charged lepton masses, the
neutrino mass-squared differences, and the mixing angles. Because of
the high dimensionality of the parameter space, the whole solution area
may be disconnected and irregular. It is difficult, if not impossible,
to find the whole solution area analytically. It is
unnecessary to find solutions which require a large cancellation among
the parameters and thus are regarded as being unlikely. 
We notice that some pioneer works use machine learning techniques to explore SUSY models, such as SUSY-AI \cite{Caron2016The} and Machine Learning Scan \cite{Ren:2017ymm}.
These machine learning techniques aim at finding all the solution ranges of the model. In this work, we are interested in finding a representative solution region, in order to see the typical prediction of the model.
 To this end,
we utilize a heuristic way to find a desirable region, it is very
efficient. Through some physical consideration, we can set a rough
initial parameter range, then the initial range is gradually
optimized until all the parameters in it are feasible. A generic
neutrino mass pattern is predicted. And the sensitiveness of each
parameter to the model results can be easily seen. Moreover, our
previous work can be also verified.

This paper is organized as follows.  In Sect. II, we will review and 
expand the discussion of the leptonic mass matrices.   Additionally, 
the physical meaning and ranges of the parameters will be also 
analyzed.  The solutions are found in Sect. III by the machine learning 
method.  Sect. IV gives discussion and prediction.  A summary is given 
in the final section.

\section{Mass analysis} 

The aim is to analyze the mass matrices of the model of Ref. 
\cite{Lei:2018lln}.  The charged lepton mass matrix is 
\begin{equation}
\label{ml}
M^l=
\begin{pmatrix}
0& \lambda_\mu v_{l_\mu}e^{i\delta_{l_\mu}}& \lambda_{\tau}v_{l\mu}e^{i\delta_{l\mu}}\\
0&\lambda_\mu v_{le}e^{i\delta_{le}}&\lambda_{\tau} v_{le}e^{i\delta_{le}}\\
0&0&y_\tau v_{d}e^{i\delta_d}\\
\end{pmatrix},
\end{equation} 
and the neutrino mass matrix is 
\begin{equation}
\label{mv}
{M}^{\nu}=-\frac{a^2}{M_{\tilde{Z}}}
\begin{pmatrix}
\lambda'_1e^{i\delta_{\lambda_1}}+v^2_{l_e}e^{2i\delta_{le}-i\delta_{Z}}&v_{l_e}v_{l_\mu}e^{i(\delta_{l_e}+\delta_{l_\mu}-i\delta_{Z})}&v_{l_e}v_{l_\tau}e^{i(\delta_{l_\tau}+\delta_{l_e})-i\delta_{Z}}\\
v_{l_e}v_{l_\mu}e^{i(\delta_{l_e}+\delta_{l_\mu}-\delta_{Z})}&\lambda'_1e^{i\delta_{\lambda_1}}+v^2_{l_\mu}e^{2i\delta_{l_\mu}-i\delta_{Z}}&v_{l_\mu}v_{l_\tau}e^{i\delta_\tau+i\delta_\mu-i\delta_{Z}}\\
v_{l_e}v_{l_\tau}e^{i\delta_\tau+i\delta_\mu-i\delta_Z}&v_{l_\mu}v_{l_\tau}e^{i(\delta_{l_\tau}+\delta_{l_\mu}-\delta_{Z})}&\lambda'_2+v^2_{l_\tau}e^{2i\delta_\tau-i\delta_{Z}}
\end{pmatrix}.
\end{equation} 
In the mass matrices, 
$\lambda'_{1,2}=\displaystyle\frac{\lambda_{1,2}\lambda_4 v_u^2M_{\tilde{Z}}}{M_T}$, 
where $M_{\tilde{Z}}$ and $M_T$ are mass parameters $\sim 10^{12}$ GeV.  
$y_\tau$, $\lambda_{\mu,\tau}$ and $\lambda_{1,2,4}$ are coupling 
constants, $v_{l_{e,\mu,\tau}}$ are the absolute vacuum expectation values 
(VEVs) of sneutrino fields, $v_{u,d}$ are the absolute VEVs of Higgs fields, 
$\delta_{Z,d,l_{e,\mu,\tau}}$are  the phases of the VEVs.  Our principle is 
that all the coupling constants are natural in the Dirac sense, namely 
they take values in the range $\sim 0.01-1$.  On the other hand, the 
naturalness in the 't Hooft sense is not required, the electroweak scale 
is due to a cancelation of large scales ($\sim 10^{12}$ GeV).  The VEVs 
are expected to vary within one order of magnitude $\sim (1-100)$ GeV, 
and their phases are naturally distributed in $(0-2\pi)$.   The 
complex matrix $M^l$ is diagonalized in the standard way.  Namely 
$M^l{M^{l\dagger}}$ is diagonalized by an unitary matrix $U_l$, 
$U_l M^l {M^l}^\dag U^\dag_l=(M^{l}_{\rm diag})^2$, where 
$M^{l}_{\rm diag}$ is written as 
\begin{equation}
\begin{pmatrix}
m_e&0&0\\
0&m_\mu&0\\
0&0&m_\tau\\
\end{pmatrix}.
\end{equation}
Analytical expressions of the charged leptons are obtained as 
\begin{equation}
m_\tau\simeq\sqrt{y^2_\tau v^2_d+\lambda^2_\tau(v^2_{l_e}+v^2_{l_\mu})}, ~~~
m_\mu\simeq\lambda_\mu\sqrt{v^2_{l_e}+v^2_{l_\mu}}, ~~~
m_e\simeq 0 \,.
\end{equation}
The complex symmetric matrix $M^\nu$ consists of two parts, 
\begin{equation}
\begin{aligned}
M^{\nu}&={M}^{\nu}_1+{M}^{\nu}_0\\
&=\frac{a^2}{M_{\tilde{Z}}}
\begin{pmatrix}
\lambda'_1e^{i\delta_{\lambda_1}}&0 & 0\\
0&\lambda'_1e^{i\delta_{\lambda_1}}& 0\\
0&0& \lambda'_2\\                                                                                                                                                                                                                                                       
\end{pmatrix}
+\frac{a^2}{M_{\tilde{Z}}e^{i\delta_Z}}
\begin{pmatrix}
v^2_{l_e}e^{2i\delta_{le}}&v_{l_e}v_{l_\mu}e^{i(\delta_{l_e}+\delta_{l_\mu})}&v_{l_e}v_{l_\tau}e^{i(\delta_{l_\tau}+\delta_{l_e})}\\
v_{l_e}v_{l_\mu}e^{i(\delta_{l_e}+\delta_{l_\mu})}&v^2_{l_\mu}e^{2i\delta_{l_\mu}}&v_{l_\mu}v_{l_\tau}e^{i\delta_\tau}\\
v_{l_e}v_{l_\tau}e^{i\delta_\tau}&v_{l_\mu}v_{l_\tau}e^{i(\delta_{l_\tau}+\delta_{l_\mu})}&v^2_{l_\tau}e^{2i\delta_\tau}
\end{pmatrix}.  
\end{aligned}
\end{equation}
Each part can have dominant contribution to neutrino masses.  First, we 
consider that $M^{\nu}_1$ is the main part, that is, the second part 
$M^{\nu}_0$ is just a correction, it is seen that the first two 
generation neutrinos are degenerate.  It is further classified into two 
cases: $\lambda'_1>\lambda'_2$ and $\lambda'_2>\lambda'_1$.  The former 
is the normal mass ordering, and the latter one inverted mass ordering.  
Second, $M^{\nu}_0$ plays the main role.   In this situation, it is 
found that $\lambda'_1$ and $\lambda'_2$ are much smaller than 
$v_{l_e}^2$.  It means the magnitude of $\lambda'_{1,2}$ should be 
taken at most $0.1$ GeV$^2$.  Thus the mass of the second generation 
neutrino is about $10^{-3}$ eV.  This is not viable and will not be 
considered.

\section{Parameter range determination}

The matrices (\ref{ml}) and (\ref{mv}) can be solved via numerical 
methods.  One way is to apply the grid search.  It divides the 
parameter space into pieces and then checks whether each piece belongs 
to the solution region.   However, the solutions are scattered in too 
many irregular or disconnected regions due to the high dimension of the 
parameter space (15D).  Finding all sets of solutions by the grid 
search is inefficient, if not unaffordable.  Rather than complex and 
fragmented solution sets, a simple and integrated range for each 
parameter is more preferred and useful in practice.  In this section, 
we make use of the heuristic search to obtain such a solution 
efficiently and effectively.

\subsection{Parameter range evaluation}
For a given candidate range, the density of the solution is usually low.  
It is reasonable if inappropriate sub-ranges are removed.  By 
estimating the solution density, we can judge a range whether it is 
good.  To avoid performing a calculation in all the chosen parameter 
ranges, which is difficult in technique, a Monte Carlo simulation can 
be constructed to estimate the whole distribution by limited samples.  
Specifically, $n$ sets of parameters are sampled randomly in given 
parameter ranges.  Then the number of solutions in sampling parameters 
obeys binomial distribution $B(n,p)$, where $p$ is the probability of 
each sampling for hitting a solution.  Finally, the expectation of $p$ 
can be obtained by Maximum Likelihood Estimation.  

\subsection{Heuristic search}
\begin{figure}[H]
\centering
\includegraphics[scale=0.6]{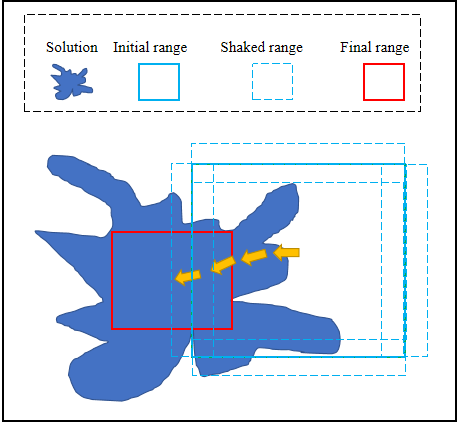}
\caption{The heuristic search}
\label{fig1}
\end{figure}

In order to find a high-density solution region efficiently, we utilize 
the heuristic search.  A reasonable initial range for every parameter 
is required, which will make the parameter range converge.  As 
Fig. \ref{fig1} shows, ranges of $m$ parameters are initialized 
manually, by shaking each side of the parameter range to increase or 
decrease the range length, $4m$ new sets of parameter ranges are 
generated.  Ranges of all the parameters are updated by those with the 
biggest expectation of $p$.  Each range is updated iteratively until 
$p\geq \theta$ (pre-set probability) or it converges.  Finally, the 
range for each parameter is output. 

\begin{figure}[H]
\centering
\centering
\includegraphics[width=12cm]{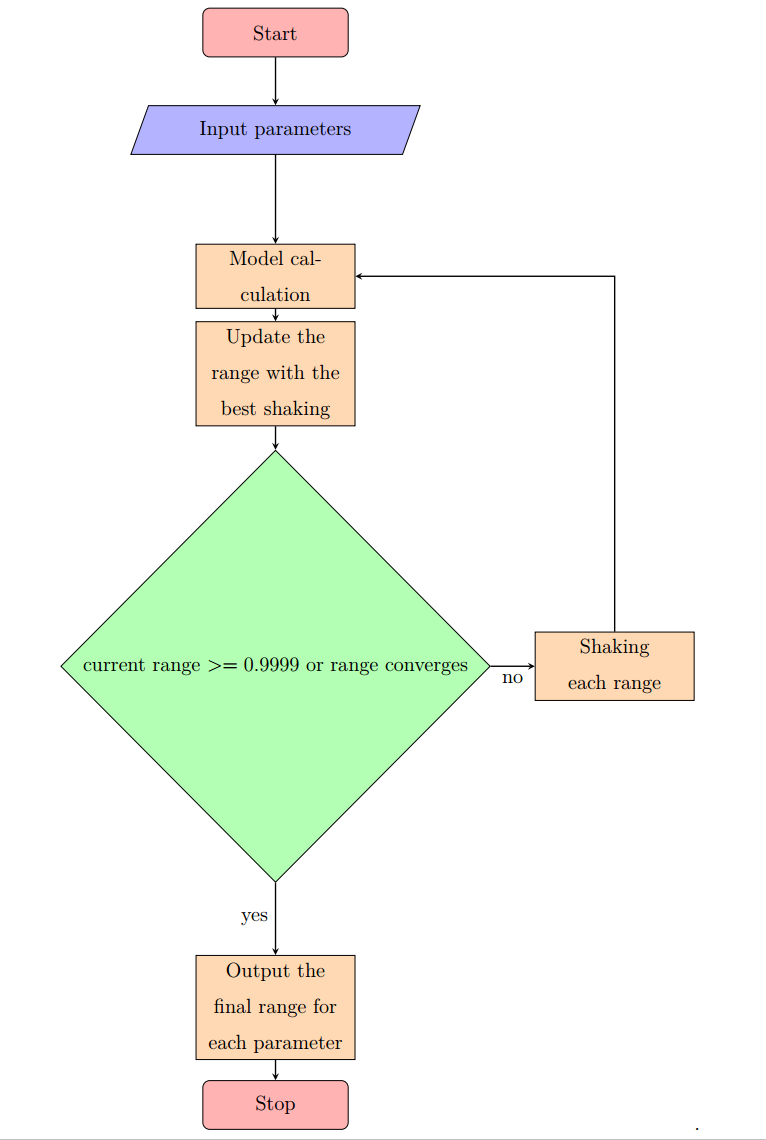}
\caption{The Majorana phases}
\label{liucheng}
\end{figure}

The heuristic method can produce and compare results by itself, the 
calculation goes on until the best parameter ranges generate almost 
identical experimental results.  Here is the method.  First, input a 
set of parameter ranges, physical quantities calculated with these 
parameters are compared to experimental values.  The probability that 
these calculated results match the experimental data can be obtained, 
it is proportional to the solution density.  When the probability is 
less than 99.99$\%$, the parameters will change 30 percent 
automatically,  and the probability will be recalculated.  Repeat the 
calculation like this until the results are almost identical to the 
experimental values with a consistency of 99.99$\%$.  The newly 
obtained parameter ranges are taken as the true ones.  Fig.\ref{liucheng} is the 
flowchart,





\begin{figure}[H]
\centering
\includegraphics[scale=0.6]{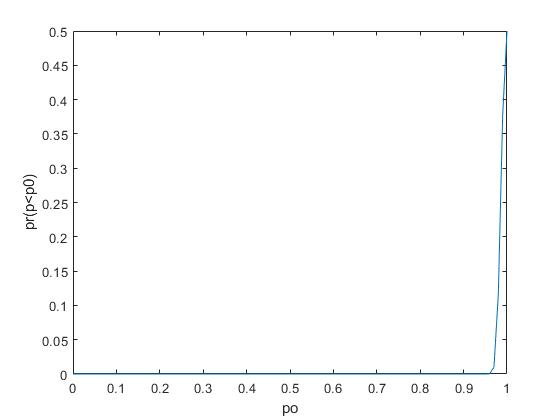}
\caption{Relation of probability $P_r(p\le p_0)$ and expectation $p_0$ when 
$N=10000$.  The abscissa $p_0$ indicates the calculated probability.} 
\label{p0} 
\end{figure}

Even if the probability of the solution in the final ranges is as high 
as $0.9999$, it does not always mean the expected value is high.  For 
instance, when we flip a fair coin,  if X denotes the value of the coin 
flip with the head, then the expected value of the random quantity X 
is $1/2$.  However, if we flip the coin ten times, it is possible to 
get the head every time.  Ten times is not an ideal number of 
experiments.  As long as we do enough many times coin-flipping 
experiments, the probability will close to the expected value.  Similar 
to that situation, it is necessary to prove that picking 10,000 points 
randomly is reasonable.  According to the Chernoff bound 
\cite{Buot2006Probability}, 
\begin{equation}
\ P_r(X\ge(1+\delta)Np)\le\frac{e^{\delta Np}}{(1+\delta)^{(1+\delta)Np}},\\
\end{equation}
namely, 
\begin{equation}
\ P_r(p\le\frac{X}{(1+\delta)N})\le\frac{e^{\delta Np}}{(1+\delta)^{(1+\delta)Np}},
\end{equation}
where function $P_r(x)$ means the probability of $x$, $N$ is the total 
times, $X$ is the number of the valid points, and 
$\displaystyle\frac{X}{(1+\delta)N}$ approximately equals to the 
calculated probability with $\delta$ standing for the uncertainty of 
the probability.  Let $p_0=\displaystyle\frac{X}{(1+\delta)N}$, $P_r(p\le p_0)$ is 
the probability that the expected value $p$ is smaller than $p_0$.  
Fig. \ref{p0} shows the relationship between $P_r(p\le p_0)$ and the 
probability $p_0$ when $N=10000$.  As can be seen from this figure that 
the probability that the expected value $p$ is smaller than 0.98 is 0 
when $N=10,000$ is chosen.  In other words, the expected value is 
almost the same as the calculated probability.  

\subsection{Experiments and results}
The parameters in the model can be divided into two parts according to 
matrices (\ref{ml}) and (\ref{mv}), their ranges can be analyzed by the 
heuristic method.  First, we consider the charged lepton sector.  For 
initial ranges, we input all the dimensionless couplings $\lambda_\tau$, 
$\lambda_\mu$, $y_\tau$ to be in (0.01-1), $v_{l_\alpha}$ in (1-10) GeV 
and $\delta_{l_e}$, $\delta_{l_\mu}$ and $\delta_d$ in $(0-2\pi)$.  
Compared with experimental data \cite{Tanabashi:2018oca}, 
$m_\mu=0.105$ GeV and $m_\tau=1.776$ GeV, the heuristic method can 
filter out the values of parameters by shaking, then the final ranges 
can be obtained. 

In the neutrino sector, the final ranges from the charged lepton sector 
are taken as initial ranges.  Moreover, other parameters $\lambda'_1$, 
$\lambda'_2$, $v_{l_\tau}$, $\delta_{Z, \lambda_1}$ and $\delta_{l_\tau}$ 
are also considered.  For the normal neutrino mass ordering, 
after multiple feedback from comparing with experimental data by 
shaking, it is found that $\lambda'_1\simeq10$ GeV$^2$, and 
$\lambda'_2\simeq 30$ GeV$^2$ as suitable initial values.  As with the 
parameters in the charged lepton sector, we enter $v_{l_\tau}$ in 
(1-10) GeV, $\delta_{l_\tau}$ and $\delta_{\lambda_1}$ in $(0-2\pi)$ as 
the initial ranges.  By repeating the shaking step, full parameter 
ranges are obtained.  Note in this sector, we filter parameter value 
ranges with experimentally known neutrino mass squared differences 
${\Delta m_{21}}^2=(7.53 \pm 0.18)\times 10^{-5}$ eV$^2$ and 
${\Delta m_{32}}^2=(2.51 \pm 0.05)\times 10^{-3}$ eV$^2$ and the mixing 
angles $\sin^2{(\theta_{12})}=0.307 \pm 0.013$, 
$\sin^2{(\theta_{23})}=0.417 \pm 0.025$ and 
$\sin^2{(\theta_{13})}=(2.12 \pm 0.08)\times 10^{-2}$ 
\cite{Tanabashi:2018oca}.   

On the other hand, for the case of the inverted neutrino mass ordering, 
by choosing $\lambda'_2>\lambda'_1$, it is found that the initial 
ranges should be $\lambda'_1\simeq (5-10)$ GeV$^2$ and 
$\lambda'_2\simeq (10-20)$ GeV$^2$.  But it always gives out empty 
final ranges.  This indicates that there is no natural solution for the 
inverted neutrino mass ordering.

\section{Analysis and prediction}

Because in the shaking step,  range boundaries change randomly, there 
are different or disconnected solutions.  Here we list three 
representative solutions in Tables \ref{tab1}, \ref{tab2} and 
\ref{tab3}.  As can be seen from Table \ref {tab1}, several 
dimensionless coefficients are almost fixed, whereas the phase 
$\delta_d$ can be chosen values almost from 0 to 2$\pi$.  This means 
the later parameters are  insensitive to the measured quantities.  It 
is seen from Tables \ref{tab2} and \ref{tab3} that $\lambda'_1$ and 
$\lambda'_2$ make major contribution to the neutrino masses. They are 
also almost fixed.

\begin{table}
\begin{tabular*}{\textwidth}{|c@{\extracolsep{\fill}}cccc|}
\hline
Parameter & Initial range & Output range 1 & Output range 2 & Output range 3\\
$\delta_{{l_e}}$  &0-$2\pi$ &  1.2566-1.5009&3.9609-4.6449&4.5752-5.1800\\                                                      
$\delta_{{l_\mu}}$ &0-$2\pi$& 5.2517-5.5449&5.2517-5.5449&2.7266-2.9156\\                                              
$\delta_{d}$  &0-$2\pi$  & 0.7540-1.4326&1.2566-3.1416&3.1416-6.2754\\
$y_{\tau}$ & 0.01- 1& 0.4525 -0.4540&0.4525-0.4550 &0.4540-0.4570     \\
$\lambda_{\tau}$ &0.01- 1&  0.4886-0.4900&0.4876-0.4890&0.4886-0.4900   \\
$\lambda_{\mu}$ &0.01- 1  &0.0600 -0.0602 &0.0598-0.0604&0.0600-0.0606\\
$\delta_{l_\tau}$ &  0-$2\pi$  &2.4288-2.500&2.4453-2.4712&2.4280-2.500\\
$\delta_{\lambda_1}$ & 0-$2\pi$  &1.5800-1.800 &1.6997-1.7396&2.7463-2.9515 \\
$\delta_{Z}$ & 0-$2\pi$  &0.0305-0.032 &0.0312-0.0323&0.0318-0.0333 \\
\hline
\end{tabular*}
\caption{The different out put of dimensionless parameter ranges}
\label{tab1}
\end{table}

\begin{table}[H]
\scriptsize
\begin{tabular*}{\textwidth}{|c@{\extracolsep{\fill}}cccc|}
\hline
Parameter & Initial range (GeV)&Output range 1(GeV) &Output range 2(GeV)&Output range 3(GeV)\\
$v_{l_e}$  & 1-10  &1.2532-1.2651&1.2585-1.2783&1.2617-1.2783\\
$v_{l_\mu}$ &1-10 & 2.5767-2.5809&2.5712-2.5770&2.5712-2.5809\\
$v_{d}$ & 1-10&2.3572-2.3661&2.3511-2.3586&2.3607-2.3661\\   
$v_{l_\tau}$ & 1-10&6.1352-6.200&6.0200-6.200&6.0788-6.0927\\
\hline
\end{tabular*}
\caption{The different output of dimensionful parameter ranges}
\label{tab2}
\end{table}

\begin{table}[H]
\scriptsize
\begin{tabular*}{\textwidth}{|c@{\extracolsep{\fill}}cccc|}
\hline
Parameter & Initial range (GeV$^2$)&Output range 1 (GeV$^2$) &Output range 2 (GeV$^2$)&Output range 3 (GeV$^2$)\\
$\lambda'_{1}$ &   10-11   &11.5800-11.700 &11.5320-11.6200&11.500-11.5720\\
$\lambda'_{2}$ &   30-31   &30.5000 -30.700&30.500-30.6200  &30.500-30.700  \\
\hline
\end{tabular*}
\caption{$\lambda'_1$ and $\lambda'_2$}
\label{tab3}
\end{table}

\begin{figure}[htbp]
\centering
\centering
\includegraphics[width=12cm]{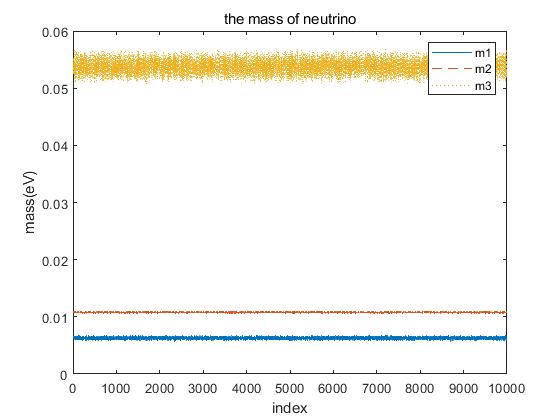}
\caption{
Neutrino masses}
\label{neutrino}
\end{figure}

With these determined parameter ranges, neutrino physical results can 
be calculated.  Firstly, three neutrino masses are predicted as in 
Fig. \ref{neutrino}.  Three generation neutrino masses are 
$m_{\nu_1}\simeq 0.007$ eV, $m_{\nu_2}\simeq 0.011$ eV and 
$m_{\nu_3}\simeq 0.051-0.056$ eV.  

.
 

Secondly, the CP-violation phase $\delta_{CP}$ is generically large.  
The area of the unitarity triangle is calculated with obtained 
parameter ranges.  Jarlskog invariant is twice of the area of the 
unitarity triangle, $
\mathcal{J}=c_{12} c_{13}^{2} c_{23} s_{12} s_{13} s_{23}\sin \delta_{CP}\equiv \mathrm {Im} (V_{\alpha i}V_{\beta j}V_{\alpha j}^{*}V_{\beta i}^{*})$.{  
$\mathcal{J}$ can be written as $
\mathcal{J}=\mathcal{J}^{max}_{CP}\sin\delta_{CP}$ \cite{Esteban:2018azc}.   By choosing $
\mathcal{J}^{max}_{CP}=0.033$,  $\sin\delta_{CP}$ is calculated.  Distribution 
of the CP violation phase is shown in Fig. \ref{CP}.  From the figure, 
$\sin\delta_{CP} \leq 0$, namely $\delta_{CP}$ takes values from $\pi$ 
to $2\pi$, and most probably $\sin\delta_{CP}\simeq -0.4$.  

Thirdly, the effective Majorana mass in neutrinoless double $\beta$ 
decays is defined as 
$|\left\langle m_{ee}\right \rangle|=|m_{\nu_1}U^2_{e1}+m_{\nu_2}U^2_{e2}+m_{\nu_3}U^2_{e3}|$.  
The effective mass is shown in Fig. \ref{eff}.  Since 10,000 points are 
taken per range, we use "index" in the abscissa in the figure to 
indicate the order of the valid points.  The ordinate indicates the 
value of the effective mass calculated at each point.  From the figure, 
it can be seen that the effective Majorana mass ranges from 
$5.5\times10^{-3}$ eV to $8.5\times10^{-3}$ eV.  The order of magnitude 
of the effective Majorana neutrino mass is about 
$\sim 7\times 10^{-3}$ eV.  

With the PMNS matrix given in the following form,  
\begin{equation}
\begin{pmatrix}
U_{e1}&U_{e2}&U_{e3}\\
U_{\mu1}&U_{\mu2}&U_{\mu3}\\
U_{\tau1}&U_{\tau2}&U_{\tau3}\\
\end{pmatrix}
\begin{pmatrix}
1&0 & 0\\
0&e^{i\rho}& 0\\
0&0& e^{i\sigma}\\                                                                                                                                                                                                                                                       
\end{pmatrix},
\end{equation}
the Majorana phases are given in Fig. \ref{majorana}.  They are 
predicted as $\rho \simeq(0.44\pi-0.63\pi)$ and 
$\sigma\simeq(0.73\pi-1.01\pi)$.  
\begin{figure}[htbp]
\centering
\includegraphics[width=12cm]{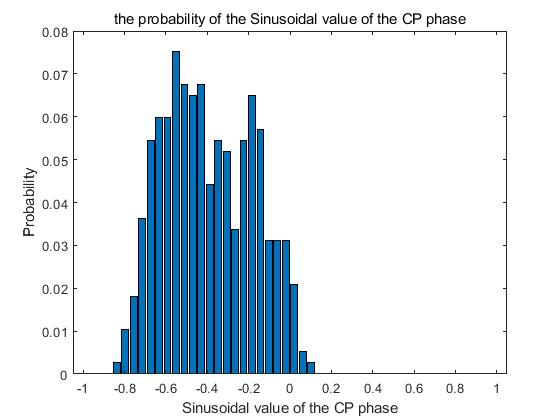}
\caption{
The distribution of the Dirac CP violation phase.}
\label{CP}
\centering
\end{figure}

\begin{figure}[H]
\centering
\centering
\includegraphics[width=12cm]{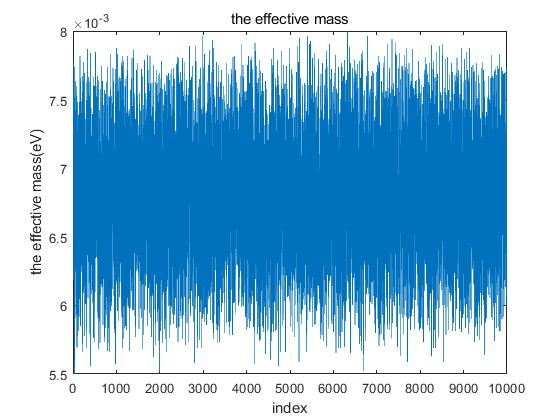}
\caption{
The range of effective Majorana neutrino mass.}
\label{eff}
\end{figure}

\begin{figure}[H]
\centering
\centering
\includegraphics[width=12cm]{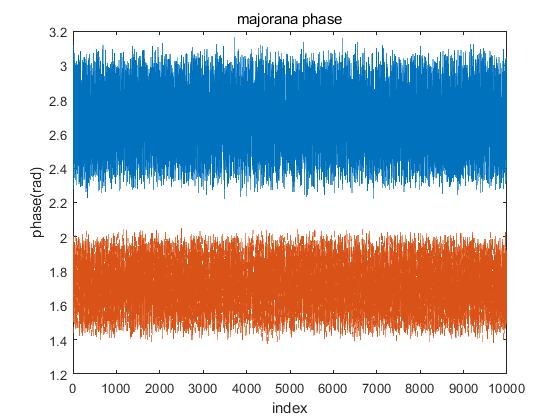}
\caption{
The Majorana phases.}
\label{majorana}
\end{figure}

It is necessary to check our last approximated anlysis of 
Ref. \cite{Lei:2018lln} by this method.  We take approximation in the 
previous work, that is, $\delta_{l\alpha}= 0$ as the input condition.  
Repeating the process of the heuristic search, neutrino masses and the 
distribution of CP violation are obtained in Figs. \ref{mass} and 
\ref{CPeg}, respectively.  It is seen from the Fig. \ref{mass} that the 
masses of the first two generation neutrinos are almost the same, 
$m_{\nu_1}$ and $m_{\nu_2}$ are around 0.02 eV. These results are 
consistent with our previous predictions.  Note, however, that the 
number of abscissa indices tells us that this solution is unlikely.  

\begin{figure}[H]
\centering
\includegraphics[width=12cm]{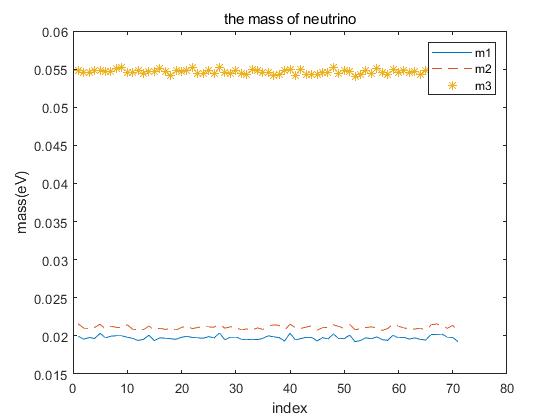}
\caption{
Neutrino masses of \cite{Lei:2018lln}}
\label{mass}
\end{figure}

\begin{figure}[htbp]
\centering
\includegraphics[width=12cm]{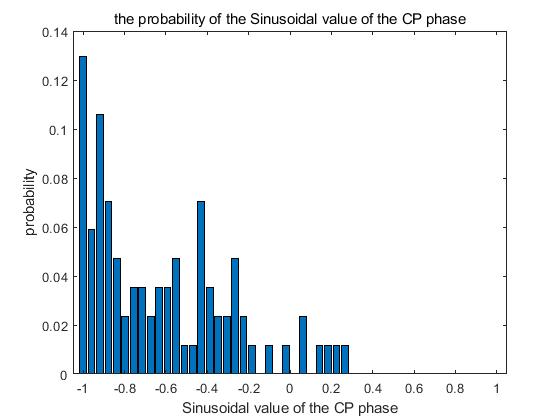}
\caption{
Distribution of the CP violation phase of Ref. \cite{Lei:2018lln}}
\label{CPeg}
\end{figure}

\section{Summary} 

We have tried to find out phenomenological consequences for neutrino 
physics of the high scale SUSY model for fermion masses.  For the lepton 
mass matrices (\ref{ml}) and (\ref{mv}) obtained from the theory model, 
the heuristic method has been applied to find the most suitable solution 
for all the physical parameters numerically.  And the Monte Carlo 
simulation has been also used to judge the reasonability of the results. 

Probabilities that can match experimental data for different ranges of 
all the parameters have been calculated out.  By constantly adjusting 
the range of parameters, we have finally found the maximum probability 
of solutions, to determine the most appropriate parameter values.  Then 
the physical quantities can be calculated.  Following results have been 
obtained.  (1) The model only supports normal hierarchical neutrino 
mass pattern.  The masses are that $m_{\nu_1}\simeq 0.007$ eV, 
$m_{\nu_2}\simeq 0.011$ eV, and $m_{\nu_3}\simeq 0.05$ eV.  (2) The 
effective Majorana neutrino mass to be discovered in neutrinoless 
double $\beta$ decay experiments, is $7\times10^{-3}$ eV.  (3) The 
Dirac CP violating phase can be anywhere from $\pi$ to $2\pi$. 

Finally, several discussions and remarks should be made.  (1) Although 
the number of parameters is a kind of many, all the basic dimensionless 
coupling constants of the model are required to be in the natural range 
($0.01-1$).  What we have pursued here is to find phenomenological 
consequences of neutrino physics due to mass matrices (\ref{ml}) and 
(\ref{mv}) resulted from a high scale supersymmetry model.  The results 
are physically meaningful.  For example, inverted neutrino mass 
ordering is not allowed with our naturalness requirement.  It is not 
trivial that right neutrino masses and mixings can be obtained without 
introducing in small parameters or accidentally large cancellation of 
the parameters.  (2) It is necessary to compare our results in this 
work with that we obtained previously with approximation 
\cite{Lei:2018lln}.  Note that what we have obtained here is the most 
probable solution.  Other solutions, like the one obtained in 
Ref. \cite{Lei:2018lln}, is not ruled out.  To be in detail, it is seen 
that the degeneracy of the first two generation neutrinos is not 
obvious compared to our previous work.  $m_{\nu_{1,2}}$ are smaller 
than that in Ref. \cite{Lei:2018lln}.  This difference is due to the 
uncertainties of the parameters.  

The effect of the phases of the sneutrino VEVs on the model is larger than we thought, especially that of $\delta_{l_\tau}$ and $\delta_{\lambda_1}$.
 Nevertheless, it is remarkable to 
note that, for each physical quantity, the result of this analysis and 
that in Ref. \cite{Lei:2018lln} is in the same order.  In terms of 
orders of magnitude, our results agree with previous ones.  Thus we 
emphasize on that this model generically has the following neutrino 
mass pattern, $m_{\nu_{1,2}}\sim 10^{-2}$ eV and 
$m_{\nu_3}\sim 5\times 10^{-2}$ eV.  
This model once predicted large $\theta_{13}$ \cite{Liu:2005qic,Liu:2012qua}. But one number is not enough for justfying a model. The results obtained in this work will be further checked in the near future. One specific feature of the results is that the first two generation neutrino masses are very close to each other.   This implies  a relatively large effective Majorana neutrino mass to be measured in neutrinoless double beta decay experiments, meanwhile with normal mass ordering. (3) Our heuristic searching method is quite efficient.
Compared to other methods, it has advantages in analyzing a given complex model with quite a lot of trigonometric function calculations, and in finding the range of dense solutions. It may have wider use in other complicated problems in particle physics.

\begin{acknowledgments}
We would like to thank Jin Min Yang and Zhen-hua Zhao for helpful 
discussions.  The authors acknowledge support from the National Natural 
Science Foundation of China (No. 11875306).
\end{acknowledgments}

\newpage

\bibliography{main}
\bibliographystyle{plain}

\end{document}